\documentclass[a4paper,11pt]{article}
\pdfoutput=1 % if your are submitting a pdflatex (i.e. if you have
\usepackage{jheppub} % for details on the use of the package, please
\usepackage[T1]{fontenc} % if needed
\usepackage{graphicx}% Include figure files
\usepackage{epsfig}  % Include figure files
\usepackage{epsf}    % Include figure files
\usepackage{dcolumn}% Align table columns on decimal point
\usepackage{bm}% bold math
\usepackage{dcolumn}% Align table columns on decimal point
\usepackage{textcomp}% Align table columns on decimal point
\usepackage{amsfonts}
\usepackage{float}
\usepackage{subfig}
\usepackage{amssymb}
\usepackage{booktabs}
\usepackage{color}
\usepackage{amsmath}
\usepackage{slashed}
\usepackage{soul}
\usepackage[]{hyperref}
  \hypersetup{
  unicode=false,          % non-Latin characters in Acrobat?s bookmarks
  pdftoolbar=true,        % show Acrobat?s toolbar?
  pdfmenubar=true,        % show Acrobat?s menu?
  pdffitwindow=true,     % window fit to page when opened
  pdfstartview={FitH},    % fits the width of the page to the window
  pdfnewwindow=true,      % links in new window
  pdfcreator={RevTeX},
  colorlinks=true,       % false: boxed links; true: colored links
  linkcolor=red,          % color of internal links
  citecolor=blue,        % color of links to bibliography
  urlcolor=blue,           % color of external links
  }
\usepackage{hypcap}

\def\beq{\begin{equation}}
\def\eeq{\end{equation}}
\def\bea{\begin{eqnarray}}
\def\eea{\end{eqnarray}}
\def\nn{\nonumber}

\def\roughly#1{\mathrel{\raise.3ex\hbox
{$#1$\kern-.75em\lower1ex\hbox{$\sim$}}}}
\def\lsim{\roughly<}
\def\gsim{\roughly>}

\def\sla#1{\raise.15ex\hbox{$/$}\kern-.57em #1}% Feynman slash
\def\ACP{A_{\rm CP}}

\def\cM{{\cal M}}
\def\cA{{\cal A}}
\def\cAt{{\tilde{\cal A}}}

\title{\boldmath CP Violation in Rare Lepton-Number-Violating $W$ Decays at the LHC}
\author[a]{Fatemeh Najafi}
\author[a]{Jacky Kumar}
\author[a]{David London}
\affiliation[a]{
{\it Physique des Particules, Universit\'e de Montr\'eal,}\\
{\it 1375 Avenue Th\'er\`ese-Lavoie-Roux, Montr\'eal, QC, Canada  H2V 0B3}
}
\emailAdd{fatemeh.najafi@umontreal.ca}
\emailAdd{jacky.kumar@umontreal.ca}
\emailAdd{london@lps.umontreal.ca}

\abstract{
	Some models of leptogenesis involve a
  quasi-degenerate pair of heavy neutrinos $N_{1,2}$ whose masses can
  be small, $O({\rm GeV})$. Such neutrinos can contribute to the rare
  lepton-number-violating (LNV) decay $W^\pm \to \ell_1^\pm \ell_2^\pm
  (q'{\bar q})^\mp$. If both $N_1$ and $N_2$ contribute, there can be
  a CP-violating rate difference between the LNV decay of a $W^-$ and
  its CP-conjugate decay. In this paper, we examine the prospects for
  measuring such a CP asymmetry $\ACP$ at the LHC. We assume a value
  for the heavy-light neutrino mixing parameter $|B_{\ell N}|^2 =
  10^{-5}$, which is allowed by the present experimental constraints,
  and consider $5~{\rm GeV} \le M_N \le 80~{\rm GeV}$. We consider
  three versions of the LHC -- HL-LHC, HE-LHC, FCC-hh -- and show that
  small values of the CP asymmetry can be measured at $3\sigma$, in
  the range $1\% \lsim \ACP \lsim 15\%$.
	}
\begin{document}
\maketitle
\flushbottom

\section{Introduction}

The standard model (SM) has been extremely successful in explaining
most of the data taken to date. Still, there are questions that remain
unanswered. For example, in the SM, neutrinos are predicted to be
massless.  However, we now know that neutrinos do have masses, albeit
very small. What is the origin of these neutrino masses? Furthermore,
are neutrinos Dirac or Majorana particles? If the latter,
lepton-number-violating (LNV) processes, such as neutrinoless
double-beta ($0\nu\beta\beta$) decay, may be observable.

The most common method of generating neutrino masses uses the seesaw
mechanism \cite{GellMann:1980vs, Yanagida, Mohapatra:1979ia}, in which
three right-handed (sterile) neutrinos $N_i$ are introduced. The
diagonalization of the mass matrix leads to three ultralight neutrinos
($m_\nu \lsim 1$ eV) and three heavy neutrinos, all of which are
Majorana.

Another question is: what is the explanation for the baryon asymmetry
of the universe? All we know is that out-of-equilibrium processes
involving baryon-number violation and CP violation are required
\cite{Sakharov:1967dj}. One idea that has been proposed to explain the
baryon asymmetry is leptogenesis. Here the idea is that CP-violating
LNV processes can produce an excess of leptons over antileptons. This
lepton asymmetry is converted into a baryon asymmetry through
sphaleron processes \cite{tHooft:1976rip, tHooft:1976snw}.

A great deal of work has been done trying to combine these two ideas.
One scenario that often arises is the appearance of a pair of heavy
neutrinos, $N_1$ and $N_2$, whose masses are nearly degenerate. With
this quasi-degenerate pair, leptogenesis can be produceed through
CP-violating decays of the heavy neutrinos \cite{Pilaftsis:1997jf,
  Pilaftsis:2003gt}, or via neutrino oscillations
\cite{Akhmedov:1998qx, Canetti:2012kh}.

One particularly intriguing aspect of this scenatio is that the
nearly-degenerate neutrinos can have masses as small as $O({\rm GeV})$
\cite{Canetti:2014dka}.  The possibility that there can be
CP-violating LNV processes involving these light sterile neutrinos has
led some authors to examine ways to see such effects in the decays of
mesons \cite{Cvetic:2013eza, Cvetic:2014nla, Dib:2014pga,
  Cvetic:2015naa, Cvetic:2015ura, Cvetic:2020lyh, Godbole:2020doo,
  Zhang:2020hwj} and $\tau$ leptons \cite{Zamora-Saa:2016ito,
  Zamora-Saa:2019naq}. Note that these studies all use as motivation
the neutrino minimal standard model ($\nu$MSM)
\cite{Appelquist:2002me, Appelquist:2003uu, Asaka:2005an,
  Asaka:2005pn}, which combines the seesaw mechanism and leptogenesis,
and even provides a candidate for dark matter.  However, it is argued
in Ref.~\cite{Drewes:2016jae} (see also Refs.~\cite{Casas:2001sr,
  Kersten:2007vk}) that the size of CP violation in the $\nu$MSM,
while large enough to explain the baryon asymmetry of the universe, is
too small to lead to a measurable effect at low energies. Still,
CP-violating effects in other models may not be so small, which is the
motivation for our work.

The idea of Refs.~\cite{Cvetic:2013eza, Cvetic:2014nla, Dib:2014pga,
  Cvetic:2015naa, Cvetic:2015ura, Cvetic:2020lyh, Godbole:2020doo,
  Zhang:2020hwj, Zamora-Saa:2016ito, Zamora-Saa:2019naq} is as
follows. The seesaw mechanism yields heavy-light neutrino mixing,
which generates a $W$-$\ell$-$N$ coupling. This leads to decays such
as $B^\pm \to D^0 \ell_1^\pm \ell_2^\pm \pi^\mp$ via $B^\pm \to D^0
W^{*\pm} (\to \ell_1^\pm N)$, with $N \to \ell_2^\pm W^{*\mp} (\to
\pi^\mp)$ \cite{Cvetic:2020lyh}. CP violation occurs because there are
two heavy neutrinos, $N = N_1$ or $N_2$, and these are nearly
degenerate in mass. The interference of the two amplitudes leads to a
difference in the rates of process and anti-process, which is a signal
of CP violation.

The key point here is that the underlying LNV process is a $W$ decay.
In the above meson and $\tau$ decays, the $W$ is virtual, but similar
effects can be searched for in the decays of real $W$s at the LHC.  To
be specific, the $0\nu\beta\beta$-like process is $W^- \to \ell_1^-
\ell_2^- (q'{\bar q})^+$. This decay has been studied extensively,
both theoretically \cite{Datta:1993nm, Ali:2001gsa, Han:2006ip,
  Chen:2013foz, Izaguirre:2015pga, Degrande:2016aje, Das:2016hof,
  Das:2017nvm, Hernandez:2018cgc} and experimentally
\cite{Abreu:1996pa, Achard:2001qv, Khachatryan:2015gha,
  Sirunyan:2018mtv, Aad:2019kiz, Aaij:2020ovh}, as a signal of LNV.
In the present paper, we push this further and study CP violation in
this decay.

We consider both the decay $W^- \to \ell_1^- \ell_2^- (q'{\bar q})^+$
and its CP-conjugate. In order to generate a CP-violating rate
difference between the two processes, the two interfering amplitudes
mediated by the nearly-degenerate $N_1$ and $N_2$ must have different
CP-odd and CP-even phases. The CP-odd phase difference is due simply
to different couplings of the two heavy neutrinos. As for the CP-even
phase difference, this can be generated through propagator effects or
heavy neutrino oscillations. (These mirror the two different ways of
producing CP-violating LNV processes for leptogenesis.) We take both
into account in our study of these decays at the LHC. We will show
that, if the new-physics parameters are such that $W^- \to \ell_1^-
\ell_2^- (q'{\bar q})^+$ is observable, a CP-violating rate asymmetry
$\ACP$ may be as well.

In Sec.~2, we consider the decay $W^- \to \ell_1^- \ell_2^- (q'{\bar
  q})^+$. We work out the individual amplitudes $\cM_i^{--}$, the
square of the total amplitude, $| \cM_1^{--} + \cM_2^{--} |^2$, and
the CP asymmetry $\ACP$. The experimental prospects for measuring
$\ACP$ are examined in Sec.~3. We compute the expected number of
events at the LHC and the corresponding minimal value of $|\ACP|$
measurable. We include the production of $W^\mp$ in $pp$ collisions,
and take into account the lifetime of the $N_i$ and experimental
efficiency. A summary \& discussion are presented in Sec.~4.

\section{\boldmath $W^- \to \ell_1^- \ell_2^- (f' {\bar f})^+$}

As described in the Introduction, the seesaw mechanism produces three
ultralight neutrinos, $\nu_j$ ($j=1,2,3$), and three heavy neutrinos,
$N_i$ ($i=1,2,3$). The flavour eigenstates $\nu_\ell$ are expressed in
terms of the mass eigenstates as follows:
\beq
\nu_\ell  = \sum_{j=1}^3 B_{\ell j} \nu_j + \sum_{i=1}^3 B_{\ell N_i} N_i ~.
\eeq
Here the parameters $B_{\ell N_i}$ describe the heavy-light neutrino
mixing. These parameters are small, but nonzero.  Because of this,
there are $W$-$\ell$-$N_i$ couplings.  We are particularly interested
in the couplings that involve the nearly-degenerate heavy neutrinos
$N_1$ and $N_2$. They are
\beq
\mathcal{L} \supset \frac{g}{\sqrt{2}} \bar \ell \gamma^\mu P_L (B_{\ell N_1} N_1+ B_{\ell N_2} N_2) W_\mu + h.c.
\eeq
These couplings generate the $W$ decay $W^- \to \ell_1^- {\bar
  N}_i$. Using the fact that the $N_i$ is Majorana ($N_i = {\bar
  N}_i$), the ${\bar N}_i$ can subsequently decay (as an $N_i$) to
$\ell_2^- W^{*+} (\to f' {\bar f})$, where $f' {\bar f} = q' {\bar q}$
or $\ell_3^+ \nu_{\ell_3}$.

This leads to the (apparently) LNV $W$ decay $W^- \to \ell_1^-
\ell_2^- (f' {\bar f})^+$. But if $f' {\bar f} = \ell_3^+
\nu_{\ell_3}$, there is a complication. The ${\bar N}_i$ can also
decay as an ${\bar N}_i$ to $\ell_3^+ W^{*-} (\to \ell_2^-
{\bar\nu}_{\ell_2}$). This is a lepton-number-conserving (LNC)
decay. But since neither the ${\bar\nu}_{\ell_2}$ nor the
$\nu_{\ell_3}$ is detected, this final state is indistinguishable from
the one above. That is, there are effectively both LNV and LNC
contributions to the same decay. Since we wish to focus on pure LNV
decays, hereafter we consider only $f' {\bar f} = q' {\bar q}$.

Thus, we have the rare LNV $W$ decay $W^- \to \ell_1^- \ell_2^- (q'
{\bar q})^+$. This is the same decay that appears with a virtual $W$
in the decays of mesons and $\tau$ leptons, studied in
Refs.~\cite{Cvetic:2013eza, Cvetic:2014nla, Dib:2014pga,
  Cvetic:2015naa, Cvetic:2015ura, Cvetic:2020lyh, Godbole:2020doo,
  Zhang:2020hwj} and \cite{Zamora-Saa:2016ito, Zamora-Saa:2019naq},
respectively. In those papers, it was pointed out that the
interference between the $N_1$ and $N_2$ contributions can lead to a
CP-violating rate difference between process and anti-process. But if
this effect is present in these processes, it should also be seen in
rare LNV decays of a real $W$. In the present paper we study the
prospects for measuring CP violation in such decays at the LHC.

\subsection{Preamble}

It is useful to make some preliminary remarks. For the decay $W^- \to
F$, where $F$ is the final state, the signal of CP violation will be a
nonzero value of
\beq
\ACP = \frac{BR(W^- \to F) - BR(W^+ \to {\bar F})}{BR(W^- \to F) + BR(W^+ \to {\bar F})} ~.
\label{ACPdef}
\eeq
Suppose this decay has two contributing amplitudes, $A$ and $B$. The
total amplitude is then
\bea
A_{\rm tot} = A + B = |A| e^{i\phi_A} e^{i\delta_A} + |B| e^{i\phi_B} e^{i\delta_B} ~,
\label{directCPamp}
\eea
where $\phi_{A,B}$ and $\delta_{A,B}$ are CP-odd and CP-even phases,
respectively. With this,
\beq
\ACP = \frac{2 |A| |B| \sin (\phi_A - \phi_B) \sin (\delta_A - \delta_B)}
     { |A|^2 + |B|^2 + 2 |A| |B| \cos (\phi_A - \phi_B) \cos (\delta_A - \delta_B)} ~.
\label{ACP_phases}
\eeq

The point is that, in order to produce a nonzero $\ACP$, the two
interfering amplitudes must have different CP-odd and CP-even phases.
In $W^- \to \ell_1^- \ell_2^- (q' {\bar q})^+$, the two amplitudes are
$W^- \to \ell_1^- {\bar N}_{1,2}$, with ${\bar N}_{1,2}$ each
subsequently decaying to $\ell_2^- (q' {\bar q})^+$. The two CP-odd
phases are $\arg[B_{\ell_1 N_1} B_{\ell_2 N_1}]$ and $\arg[B_{\ell_1
    N_2} B_{\ell_2 N_2}]$, which can clearly be different.

For the CP-even phases, things are a bit more complicated. The usual
way such phases are generated is via gluon exchange (which is why they
are often referred to as ``strong phases''). However, since this decay
involves the $W^\pm$, $\ell_i^\pm$ and $N_i$, which are all
colourless, this is not possible. Instead, the CP-even phases can be
generated in one of two ways. First, the propagator for the $N_i$ is
proportional to
\bea
\frac{1}{(p_N^2 - M_{N_i}^2) + i M_{N_i} \Gamma_{N_i}}
&=& \frac{1}{\sqrt{(p_N^2 - M_{N_i}^2)^2 + M_{N_i}^2 \Gamma_{N_i}^2}} \, e^{i \eta_i} ~, \nn\\
{\rm with} ~~~~~
\tan\eta_i &=& \frac{- M_{N_i} \Gamma_{N_i}}{(p_N^2 - M_{N_i}^2)} ~.
\label{propagator}
\eea
Thus, $\eta_i$ is the CP-even phase associated with the propagator.
Since $N_1$ and $N_2$ do not have the same mass -- they are nearly,
but not exactly, degenerate -- if one of the $N_i$ is on shell ($p_N^2
= M_{N_i}^2$), the other is not. This creates a nonzero CP-even phase
difference: the on-shell $N_i$ has $\eta_i = - \pi/2$, while the
CP-even phase of the other $N_i$ obeys $|\eta_i| < \pi/2$. This leads
to what is known as resonant CP violation\footnote{Note that it is
  important that the $N_i$ be nearly degenerate. From
  Eq.~(\ref{ACP_phases}) we see that $\ACP$ is sizeable only when the
  two contributing amplitudes are of similar size ($|A| \sim
  |B|$). But if the masses of $N_1$ and $N_2$ were very different, the
  size of their contributions to the decay would also be very
  different, leading to a small $\ACP$.}.

A second way of generating a CP-even phase difference is through
oscillations of the heavy neutrinos. As we will see below, the time
evolution of a heavy $N_i$ mass eigenstate involves $e^{-i E_i t}$ (in
addition to the exponential decay factor). Since $N_1$ and $N_2$ do
not have tne same mass, their energies are different, leading to
different $e^{-i E_i t}$ factors. This is another type of CP-even phase
difference, and can also lead to CP violation.

Below we derive the amplitudes for $W^- \to \ell_1^- {\bar N}_i$, with
each ${\bar N}_i$ subsequently decaying to $\ell_2^- (q' {\bar q})^+$,
including both types of CP-even phases.

\subsection{\boldmath Decay amplitudes $\cM_i^{--}$}

\begin{figure}[!htbp]
\begin{center}
 \includegraphics[clip, trim=9cm 20cm 5cm 3.5cm, width=0.60\textwidth]{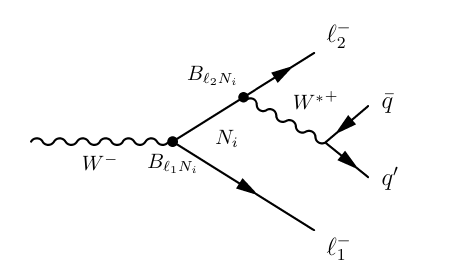}
\end{center}
\caption{\small Diagram for $W^- \to \ell_1^- \ell_2^- (q'{\bar q})^+$
  via an intermediate $N_i$. There is no arrow on the $N_i$ line
  because it is a Majorana particle and the decay is fermion-number
  violating.}
\label{Ndecayfig}
\end{figure}

Consider the diagram of Fig.~\ref{Ndecayfig}, with $N_i = N_1$. If
this were the only contribution, its amplitude could be written simply
as the product of two amplitudes, one for $W^- \to \ell_1^- {\bar
  N}_i$, the other for ${\bar N}_1 \to \ell_2^- (q' {\bar q})^+$.
However, because there are contributions from $N_1$ and $N_2$, and
because $N_1$ and $N_2$ cannot be on shell simultaneously, we must
include the heavy neutrino propagator.

Furthermore, although the neutrino is produced as ${\bar N}_i$, it
actually decays as $N_i$, leading to the fermion-number-violating and
LNV process $W^- \to \ell_1^- \ell_2^- (q'{\bar q})^+$. This implies
that (i) conjugate fields will be involved in the amplitudes, and (ii)
the amplitudes will be proportional to the neutrino mass.

We can now construct the amplitudes $\cM_i^{--} \equiv \cM(W^- \to
\ell_1^- {\bar N}_i, N_i \to \ell_2^- W^{*+} (\to (q' {\bar q})^+)$.
Writing $\cM_i^{--} = \cM_i^{\mu\nu} \epsilon_\mu j_\nu$, where
$\epsilon_\mu$ is the polarization of the initial $W^-$ and $j_\nu =
\frac{g}{\sqrt 2} {\bar q} \gamma_\nu P_L q'$ is the current of
final-state particles to which $W^{*+}$ decays, we have
\bea
\cM_i^{\mu\nu} &=&
\bar \ell_1 \gamma^\mu P_L \left (\frac{g}{\sqrt 2} B_{\ell_1 N_i} \right ) N_i
\times e^{-\Gamma_i t/2} e^{-iE_i t} \times
\bar \ell_2 \gamma^\nu P_L \left (\frac{g}{\sqrt 2} B_{\ell_2 N_i} \right ) N_i \nn\\
&=& \frac{g^2}{2} B_{\ell_1 N_i} B_{\ell_2 N_i} \,
\bar \ell_1 \gamma^\mu P_L N_i \,
\overline{N_i^c} \gamma^\nu P_R \ell_2^c \times e^{-\Gamma_i t/2} e^{-iE_i t} \nn\\
&\to&
\frac{g^2}{2} B_{\ell_1 N_i} B_{\ell_2 N_i} \,
\bar \ell_1 \gamma^\mu P_L
\frac{\sla{p} + M_i}{p_N^2 - M^2_i + i \Gamma_i M_i} \,
\gamma^\nu P_R \ell_2^c \times e^{-\Gamma_i t/2} e^{-iE_i t} \nn\\
&=&
\frac{ \frac{g^2}{2} B_{\ell_1 N_i} B_{\ell_2 N_i} \, M_i \, e^{-\Gamma_i t/2} e^{-iE_i t}}{p_N^2 - M^2_i + i \Gamma_i M_i} \,
L^{\mu\nu} ~,
\eea
where $L^{\mu\nu} = \bar \ell_1 \gamma^\mu \gamma^\nu P_R \ell_2^c$.
In the first line, the first term is the amplitude for $W^- \to
\ell_1^- {\bar N}_i$, the second term is the time dependence of the
$N_i$ state, and the third term is the amplitude for $N_i \to \ell_2^-
W^{*+}$. The $e^{-iE_i t}$ factor is due to the quantum-mechanical
evolution of the $N_i$ state; its energy $E_i$ is evaluated in the
rest frame of the decaying $W$. In the second line, we have taken the
transpose of the third term, writing the current in terms of conjugate
fields, $\psi^c = C{\bar\psi}^T$. And in the third line, we have
replaced $N_i \, \overline{N_i^c}$ by the neutrino propagator.

Another contribution to this process comes from a diagram like that of
Fig.~\ref{Ndecayfig}, but with $\ell_1 \leftrightarrow \ell_2$. The
amplitude for this diagram is the same as that above, but with (i)
$p_N \to p'_N$ and (ii) $\bar \ell_1 \gamma^\mu \gamma^\nu P_R
\ell_2^c \to \bar \ell_2 \gamma^\mu \gamma^\nu P_R \ell_1^c = -\bar
\ell_1 \gamma^\nu \gamma^\mu P_R \ell_2^c$. Now, if $\ell_1 \ne
\ell_2$, one simply adds the diagrams, while if $\ell_1 = \ell_2$,
there is an additional minus sign. Thus, the amplitude for this second
diagram is
\bea
\cM_i^{\prime\mu\nu} &=& \pm \frac{ \frac{g^2}{2} B_{\ell_1 N_i} B_{\ell_2 N_i} \, M_i \, e^{-\Gamma_i t/2} e^{-iE_i t}}
   {{p'_N}^2 - M^2_i + i \Gamma_i M_i} \, L^{\mu\nu} ~,
\eea
and the total amplitude is $\cM_i^{\mu\nu} + \cM_i^{\prime\mu\nu}$.
Now, the dominant contributions to these amplitudes come from (almost)
on-shell $N_i$s. This means that, while both diagrams lead to $W^- \to
\ell_1^- \ell_2^- (q' {\bar q})^+$, the final-state particles do not
have the same momenta in the two cases. As a result, when the total
amplitude is squared, $\cM_i^{\mu\nu}$ and $\cM_i^{\prime\mu\nu}$ will
not interfere. Thus, we can consider only the $\cM_i^{\mu\nu}$; the
contribution from the $\cM_i^{\prime\mu\nu}$ will be identical.

We can now compute $\cM^{\mu\nu} = \cM_1^{\mu\nu} + \cM_2^{\mu\nu}$.
Writing
\beq
B_{\ell_1 N_1} B_{\ell_2 N_1} = B_1 e^{i \phi_1} ~~,~~~~
B_{\ell_1 N_2} B_{\ell_2 N_2} = B_2 e^{i \phi_2} ~,
\eeq
we have
\beq
\cM^{\mu\nu} = \frac{g^2}{2} \left(
  \frac{M_1 B_1 e^{i \phi_1} e^{-\Gamma_1 t/2} e^{-iE_1 t}}{p_N^2-M_1^2+i\Gamma_1 M_1}
+ \frac{M_2 B_2 e^{i \phi_2} e^{-\Gamma_2 t/2} e^{-iE_2 t}}{p_N^2-M_2^2+i\Gamma_2 M_2} \right) L^{\mu\nu} ~.
\label{Wdecayamp}
\eeq

\subsection{\boldmath $|\cM_{\rm tot}^{--}|^2$}

The complete amplitude is $\cM_{\rm tot}^{--} = \cM^{\mu\nu} \epsilon_\mu
j_\nu = (g^2/2) \, \cA_{--}(t) \, L^{\mu\nu}
\, \epsilon_\mu j_\nu$, where $\cA_{--}(t)$ is the piece in parentheses
in Eq.~(\ref{Wdecayamp}).  The next step is to compute $|\cM_{\rm
  tot}^{--}|^2$.

From the point of view of studying CP violation in the decay $W^- \to
\ell_1^- \ell_2^- (q'{\bar q})^+$, the most important term in
$\cM_{\rm tot}^{--}$ is $\cA_{--}(t)$. It is instructive to compare this
quantity with Eq.~(\ref{directCPamp}) above. In the first term of
$\cA_{--}(t)$, we can identify the CP-odd phase ($\phi_1$) and the
CP-even phase associated with neutrino oscillations ($-E_1 t$). There
is also a (different) CP-even phase $\eta_i$ associated with the
propagator [see Eq.~(\ref{propagator})]. The phases of the second term
can be similarly identified.

Consider now $|\cA_{--}(t)|^2$. We have
\bea
|\cA_{--}(t)|^2 &=&
\frac{M_1^2  B_1^2 e^{-\Gamma_1 t}}{(p_N^2-M_1^2)^2+\Gamma_1^2 M_1^2}+
\frac{M_2^2  B_2^2 e^{-\Gamma_2 t}}{(p_N^2-M_2^2)^2+\Gamma_2^2 M_2^2}  \nn\\
&& \hskip5truemm +~2 {\rm Re}
\left( \frac{M_1 M_2 B_1 B_2 \, e^{- i \delta \phi} \, e^{-\Gamma_{\rm avg}t } \, e^{-i \Delta E t}}
     {(p_N^2-M_1^2+i \Gamma_1 M_1)(p_N^2-M_2^2 - i \Gamma_2 M_2)}   \right) ~,
\label{A2}
\eea
where
\beq
\Gamma_{\rm avg}= \frac12 (\Gamma_1+\Gamma_2) ~,~~ \Delta E \equiv E_1- E_2 =
\frac{M_1^2-M_2^2}{2 M_W} ~,~~ \delta \phi \equiv \phi_2 - \phi_1 ~.
\label{DeltaEdef}
\eeq
There are two simplifications that can be made. First, in order to
compute the rate for the decay, it will be necessary to integrate over
the phase space of the final-state particles. Due to energy-momentum
conservation, this will involve an integral over $p_N$. Since the
$N_i$ can go on shell, we can use the narrow-width approximation to
replace
\beq
\frac{1}{(p_N^2-M_i^2)^2+\Gamma_i^2 M_i^2} ~~ \to ~~ \frac{\pi}{\Gamma_i M_i} \, \delta(p_N^2-M_i^2) ~.
\eeq
Second, although it is important to take neutrino oscillations into
account in considerations of CP violation, we do not focus on actually
measuring such oscillations. (This is examined in
Refs.~\cite{Antusch:2017ebe, Cvetic:2018elt, Cvetic:2019rms}.)  That is, we can
integrate over time: $\int_0^\infty {dt} |\cA_{--}(t)|^2 = |{\cal
  A}_{--}|^2$. Note that, in integrating to $\infty$, we assume that the
$N_i$ are heavy enough that their lifetimes are sufficiently small
that most $N_i$s decay in the detector. We will quantify this in the
next Section.

Now consider the interference term. Using the narrow-width
approximation, the product of propagators can be written
\bea
&& \frac{1}{(p_N^2-M_1^2 +i \Gamma_1 M_1)(p_N^2-M_2^2 - i \Gamma_2 M_2)} = \nn\\
&& \hskip3truecm
\frac{\Gamma_1 M_1 \pi \delta(p_N^2-M_2^2)}{(\Delta M^2)^2 + \Gamma_1^2 M_1^2}
+ \frac{\Gamma_2 M_2 \pi \delta(p_N^2-M_1^2)}{(\Delta M^2)^2 + \Gamma_2^2 M_2^2} \nn\\
&& \hskip3truecm -~\frac{i\Delta M^2 \pi \delta(p_N^2-M_2^2)}{(\Delta M^2)^2+\Gamma_1^2 M_1^2}
- \frac{i\Delta M^2 \pi \delta(p_N^2-M_1^2)}{(\Delta M^2)^2+\Gamma_2^2 M_2^2} ~,
\label{propagatorphase}
\eea
where $\Delta M^2 \equiv M_1^2 -M_2^2$. Note that the imaginary part
is proportional to $\Delta M^2 = (M_1 - M_2)(M_1 + M_2) \equiv \Delta
M (M_1 + M_2)$. Referring to Eq.~(\ref{propagator}), we see that the
CP-even phase difference $\eta_1 - \eta_2$ is proportional to $\Delta
M$.

Putting all the pieces together, we obtain
\bea
\label{A--_squared}
|\cA_{--}|^2 &=&
\frac{\pi M_1 B_1^2}{ \Gamma_1^2} \delta(p_N^2-M_1^2) + \frac{\pi M_2 B_2^2}{ \Gamma_2^2} \delta(p_N^2-M_2^2) \\
&& \hskip-15truemm +~\frac{2 M_1 M_2 B_1 B_2}{\Gamma^2_{\rm avg} +(\Delta E)^2} \,\
\left ( \frac{\Gamma_1 M_1 \pi \delta(p_N^2-M_2^2)}{(\Delta M^2)^2 + \Gamma_1^2 M_1^2}
+ \frac{\Gamma_2 M_2 \pi \delta(p_N^2-M_1^2)}{(\Delta M^2)^2 + \Gamma_2^2 M_2^2} \right )
(\cos(\delta \phi) \Gamma_{\rm avg} - \Delta E \sin (\delta \phi)) \nn\\
&& \hskip-15truemm +~\frac{2 M_1 M_2 B_1 B_2}{\Gamma^2_{\rm avg} +(\Delta E)^2} \,\
 \left ( \frac{\Delta M^2 \pi \delta(p_N^2-M_2^2)}{(\Delta M^2)^2+\Gamma_1^2 M_1^2}
+ \frac{\Delta M^2 \pi \delta(p_N^2-M_1^2)}{(\Delta M^2)^2+\Gamma_2^2 M_2^2}  \right )
(\cos (\delta \phi )\Delta E + \sin(\delta \phi) \Gamma_{\rm avg} ) ~. \nn
\eea

\subsection{\boldmath CP violation}

The time-integrated square of the amplitude for $W^- \to \ell_1^-
\ell_2^- (q'{\bar q})^+$ is therefore $|\cM_{--}|^2 = (g^2/2)^2 \,
|\cA_{--}|^2 \, |L^{\mu\nu} \, \epsilon_\mu j_\nu|^2$.  The CP
asymmetry is defined as [see Eq.~(\ref{ACPdef})]
\beq
\ACP = \frac{\int {d \rho} \, ( |\cM_{--}|^2 - |\cM_{++}|^2)}
     {\int {d \rho} \, (|\cM_{--}|^2 + |\cM_{++}|^2)}
     = \frac{\int {d \rho} \, ( |\cA_{--}|^2 - |\cA_{++}|^2)|L^{\mu\nu} \, \epsilon_\mu j_\nu|^2}
     {\int {d \rho} \, (|\cA_{--}|^2 + |\cA_{++}|^2)|L^{\mu\nu} \, \epsilon_\mu j_\nu|^2} ~,
\label{ACPBRdef}
\eeq
where $|\cA_{++}|^2$ is obtained from $|\cA_{--}|^2$
[Eq.~(\ref{A--_squared})] by changing the sign of the CP-odd phase,
and $\int d \rho$ indicates integration over the phase space.

For the phase-space integration, the only pieces that depend on the
integration variables are the delta function $\delta(p_N^2-M_i^2)$ in
Eq.~(\ref{A--_squared}) and $|L^{\mu\nu} \, \epsilon_\mu
j_\nu|^2$. The phase-space integrals are therefore
\beq
{\cal I}(M_i) = \int {d \rho} \, \pi \delta(p_N^2-M_i^2) |L^{\mu\nu} \, \epsilon_\mu j_\nu|^2 ~.
\eeq
In Ref.~\cite{Dib:2014pga}, it was shown that, since $M_1 \simeq M_2$,
${\cal I}(M_1) \simeq {\cal I}(M_2)$. Thus, to a very good
approximation, these terms cancel in Eq.~(\ref{ACPBRdef}), so that
\beq
\ACP = \frac{|\cAt_{--}|^2 - |\cAt_{++}|^2}{|\cAt_{--}|^2 + |\cAt_{++}|^2} ~,
\label{ACPafterint}
\eeq
where $\cAt_{--}$ ($\cAt_{++}$ ) is the same as $\cA_{--}$
($\cA_{++}$), but with the $\pi \delta(p_N^2-M_i^2)$ factors removed.

In the numerator we have
\bea
&& |\cAt_{--}|^2 - |\cAt_{++}|^2 = \\
&& \hskip0.5truecm  -~\frac{2 M_1 M_2 B_1 B_2}{\Gamma^2_{\rm avg} +(\Delta E)^2} \,\ 
\left ( \frac{\Gamma_1 M_1  }{(\Delta M^2)^2 + \Gamma_1^2 M_1^2}
+ \frac{\Gamma_2 M_2  }{(\Delta M^2)^2 + \Gamma_2^2 M_2^2} \right )
(2 \Delta E \sin (\delta \phi)) \nn\\
&& \hskip0.5truecm +~\frac{2 M_1 M_2 B_1 B_2}{\Gamma^2_{\rm avg} +(\Delta E)^2} \,\
 \left ( \frac{\Delta M^2  }{(\Delta M^2)^2+\Gamma_1^2 M_1^2} 
+ \frac{\Delta M^2  }{(\Delta M^2)^2+\Gamma_2^2 M_2^2}  \right ) 
(2 \sin(\delta \phi) \Gamma_{\rm avg} ) ~. \nn
\eea
In Eq.~(\ref{ACP_phases}), we see that $\ACP$ is proportional to $\sin
(\phi_A - \phi_B) \sin (\delta_A - \delta_B)$, i.e., a nonzero $\ACP$
requires that the two interfering amplitudes have different CP-odd and
CP-even phases. This is also true in the present case. Above, both
terms are proportional to $\sin (\delta \phi)$ ($\delta\phi$ is the
CP-odd phase difference). In the first term, the CP-even phase arises
due to neutrino oscillations: $\sin (\delta_A - \delta_B)$ is
proportional to $\Delta E$. And in the second term, the CP-even phase
difference comes from the propagators [see
  Eq.~(\ref{propagatorphase})]: it is proportional to $\Delta M$.  In
the denominator,
\bea
&& |\cAt_{--}|^2 + |\cAt_{++}|^2 =
\frac{2 M_1 B_1^2}{ \Gamma_1^2}  + 
\frac{2 M_2 B_2^2}{ \Gamma_2^2}  \\
&& \hskip0.5truecm +~\frac{2 M_1 M_2 B_1 B_2}{\Gamma^2_{\rm avg} +(\Delta E)^2} \,\ 
\left ( \frac{\Gamma_1 M_1  }{(\Delta M^2)^2 + \Gamma_1^2 M_1^2}
+ \frac{\Gamma_2 M_2  }{(\Delta M^2)^2 + \Gamma_2^2 M_2^2} \right )
(2\cos(\delta \phi) \Gamma_{\rm avg} ) \nn\\
&& \hskip0.5truecm -~\frac{2 M_1 M_2 B_1 B_2}{\Gamma^2_{\rm avg} +(\Delta E)^2} \,\
 \left ( \frac{\Delta M^2  }{(\Delta M^2)^2+\Gamma_1^2 M_1^2} 
+ \frac{\Delta M^2  }{(\Delta M^2)^2+\Gamma_2^2 M_2^2}  \right ) 
(2\cos (\delta \phi )\Delta E  ) ~. \nn
\eea

We now make the (reasonable) approximations that $\Gamma_1 \simeq
\Gamma_2 \equiv \Gamma$ and $M_1 \simeq M_2 \equiv M_N$ (but $\Delta M
\ne 0$ and is $\ll M_N$). With the assumption that $B_1 = B_2$, $\ACP$
takes a simple form:
\beq
\ACP = \frac{ 2 (2 y - x) \sin \delta \phi}{ ( 1+x^2 ) ( 1 + 4y^2 ) + 2 ( 1-2xy ) \cos \delta \phi } ~,
\label{ACPsimple}
\eeq
where
\beq
x \equiv \frac{\Delta E}{\Gamma} ~~,~~~~ y \equiv \frac{\Delta M}{\Gamma} ~.
\label{xydefs}
\eeq

Once again comparing to Eq.~(\ref{ACP_phases}), we see that $x$ and
$y$ each play the role of the CP-even phase-difference term $\sin
(\delta_A - \delta_B)$. Now, $x$ and $y$ reflect CP-even phases
arising from neutrino oscillations and the neutrino propagator,
respectively.  However, they are not, in fact, independent. From
Eq.~(\ref{DeltaEdef}), we have
\beq
\Delta E = \frac{M_1^2-M_2^2}{2 M_W} = \frac{2 \, \Delta M \, M_N}{2 M_W}
~~~~ \Longrightarrow ~~~~ x = y \, \frac{M_N}{M_W} ~.
\eeq
Thus, $y$ is always present; $x$ is generally subdominant, except for
large values of $M_N$.

Furthermore, we note that $x$ and $y$ have the same sign, and that
$|x| < |y|$. Thus, $|2y - x| \le |2y|$. That is, as $|x|$ increases,
$\ACP$ decreases. We therefore expect to see smaller CP-violating
effects for larger values of $M_N$. The reason this occurs is as
follows. Above, we said that $x$ and $y$ each play the role of $\sin
(\delta_A - \delta_B)$. However, in this system, their contributions
have the opposite sign, hence the factor $2 y - x$ in
Eq.~(\ref{ACPsimple}).

\begin{figure}[!htbp]
\begin{center}
\includegraphics[clip, trim=0.3cm 0cm 0.2cm 0.2cm, width=0.55\textwidth]{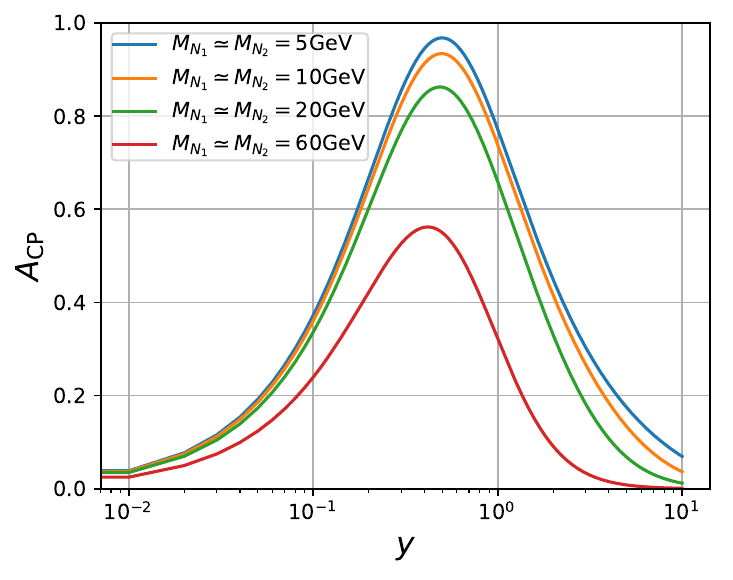}
\end{center}
\caption{\small Value of $\ACP$ as a function of $y$, for $\delta \phi
  = \pi/2$ and for various values of $M_N$. For negative values of
  $y$, $\ACP \to -\ACP$.}
\label{ACPfig}
\end{figure}

In order to get an estimate of the potential size of $\ACP$, we set
$\delta \phi = \pi/2$. In Fig.~\ref{ACPfig}, we show $\ACP$ as a
function of $y$, for various values of $M_N$. We see that large values
($\ge 0.9$) of $|\ACP|$ can be produced for light $M_N$. The maximal
values of $|\ACP|$ are found when $y \simeq \pm\frac12$, with $|\ACP|$
decreasing for larger values of $|y|$. As expected, the size of
$|\ACP|$ decreases as $M_N$ increases, with $|\ACP|_{\rm max} < 0.6$
for larger values of $M_N$.

\section{Experimental Analysis}

In this section, we explore the prospects for measuring $\ACP$ at the
LHC. We consider three versions of the LHC: (i) the high-luminosity
LHC (HL-LHC, $\sqrt{s} = 14$ TeV, peak ${\cal L}_{\rm int} = 3~{\rm
  ab}^{-1}$), (ii) the high-energy LHC (HE-LHC, $\sqrt{s} = 27$ TeV,
peak ${\cal L}_{\rm int} = 15~{\rm ab}^{-1}$)
\cite{Zimmermann:2017bbr}, (iii) the future circular
collider\footnote{The Future $e^+ e^-$ Circular Collider, FCC-ee
  (TLEP) \cite{Blondel:2014bra}, would also be a promising place to
  make this measurement.}  (FCC-hh, $\sqrt{s} = 100$ TeV, peak ${\cal
  L}_{\rm int} = 30~{\rm ab}^{-1}$) \cite{Golling:2016mxw}. We
implement the model in {\tt FeynRules} \cite{Degrande:2016aje,
  Alloul:2013bka} and use {\tt MadGraph} \cite{Alwall:2014hca} to
generate events.

The CP asymmetry of Eq.~(\ref{ACPafterint}) involves the branching
ratios of the decay $W^- \to \ell_1^- \ell_2^- (q'{\bar q})^+$ and its
CP-conjugate decay, $W^+ \to \ell_1^+ \ell_2^+ (q{\bar q}')^-$.
Another way of describing $\ACP$ is: given an equal number of initial
$W^-$ and $W^+$ bosons,
\beq
\ACP = \frac{N_{--} - N_{++}}{N_{--} + N_{++}} ~,
\label{ACPNumbers}
\eeq
where $N_{--}$ and $N_{++}$ are the number of observed events of $W^-
\to \ell_1^- \ell_2^- (q'{\bar q})^+$ and $W^+ \to \ell_1^+ \ell_2^+
(q{\bar q}')^-$, respectively.

But there is a problem: these decays are not measured directly at the
LHC.  Instead, one has $pp$ collisions, so that the processes are
${\bar u}_i d_j \to W^- \to \ell_1^- \ell_2^- (q'{\bar q})^+$ and
${\bar d}_j u_i \to W^+ \to \ell_1^+ \ell_2^+ ({\bar q}'q)^-$, where
$u_i$ and $d_j$ represent up-type and down-type quarks, respectively.
Since protons do not contain equal amounts of ${\bar u}_i d_j$ and
${\bar d}_j u_i$ pairs, the number of $W^-$ and $W^+$ bosons produced
will not be the same, and this must be taken into account in the
definition of the CP asymmetry.

This is done by changing Eq.~(\ref{ACPNumbers}) to
\beq
\ACP = \frac{ N_{--}^{pp}/\sigma^- - N_{++}^{pp}/\sigma^+}{ N_{--}^{pp}/\sigma^- + N_{++}^{pp}/\sigma^+}
= \frac{ R_W N_{--}^{pp} - N_{++}^{pp}}{ R_W N_{--}^{pp} + N_{++}^{pp}} ~,
\eeq
where $N_{--}^{pp}$ and $N_{++}^{pp}$ are the number of observed
events of $pp \to X W^- (\to \ell_1^- \ell_2^- (q'{\bar q})^+)$ and
$pp \to X W^+ (\to \ell_1^+ \ell_2^+ ({\bar q}' q)^-)$, respectively,
and $R_W = \sigma^+/\sigma^-$, with
\beq
\sigma^+ = \sigma(pp \to W^+ X) ~~,~~~~ \sigma^- = \sigma(pp \to W^- X) ~.
\eeq
Experimentally, it is found that $R_W = 1.295 \pm 0.003~(stat) \pm
0.010~(syst)$ at $\sqrt{s} = 13$ TeV \cite{Aad:2016naf}. Presumably,
$R_W$ can be measured with equally good precision (if not better) at
higher energies, so it is clear how to obtain a CP-violating
observable from the experimental measurements\footnote{In
  Ref.~\cite{Liu:2019qfa}, it is argued that a more promising way to
  search for $W^- \to \ell_1^- \ell_2^- (q'{\bar q})^+$ is to use
  $W^-s$ coming from the decay of a ${\bar t}$. If this is true, then
  if such a decay is observed, one can measure CP violation in these
  decays using the above formalism. And since top quarks mainly arise
  through $t{\bar t}$ production, there are equal numbers of $W^-$ and
  $W^+$ bosons, so that an adjustment using $R_W$ is not required.}.

%With {\tt MadGraph}, we find that this ratio does not change much at
%higher energies: $R_W = 1.39$ (13 TeV), 1.38 (14 TeV), 1.31 (27 TeV),
%1.22 (100 TeV). {\bf DL: what do we do about the fact that MadGraph
%disagrees with experiment by so much?}

Now, given a CP asymmetry $\ACP$, the number of events ($N_{\rm
  events} = N_{--}^{pp} + N_{++}^{pp}$) required to show that it is
nonzero at $n\sigma$ is
\beq
N_{\rm events} = \frac{n^2}{A_{CP}^2 \, \epsilon} ~,
\eeq
where $\epsilon$ is the experimental efficiency. This can be turned
around to answer the question: given a certain total number of events
$N_{\rm events}$, what is the smallest value of $|\ACP|$ that can be
measured at $n\sigma$?

There are two ingredients to establishing $N_{\rm events}$. The first
is the cross section for $pp \to X W^\mp$, multiplied by the branching
ratio for $W^\mp \to \ell_1^\mp {\bar N}_i (N_i)$, and further
multiplied by the branching ratio for the decay of ${\bar N}_i (N_i)$
to the final state of interest. The branching ratio for $W^\mp \to
\ell_1^\mp {\bar N}_i (N_i)$ depends on the value of the heavy-light
mixing parameter $|B_{\ell_1 N_1}|^2$. Constraints on this quantity
can be obtained from experimental searches for the same
$0\nu\beta\beta$-like process we consider here. A summary of these
constraints can be found in Ref.~\cite{Deppisch:2015qwa}. For 5 GeV
$\lsim M_N \lsim$ 50 GeV, $|B_{\ell N}|^2 \lsim 10^{-5}$ ($\ell = e,
\mu, \tau$), but the constraint is weaker for larger values of $M_N$.
In our analysis, to be conservative, we take $|B_{\ell N}|^2 =
10^{-5}$ for all values of $M_N$.

We now use {\tt MadGraph} to calculate the cross sections for $pp \to
X \ell_1^- {\bar N}$, with ${\bar N} \to \ell_2^- (q'{\bar q})^+$ and
$pp \to X \ell_1^+ N$, with $N \to \ell_2^+ ({\bar q}'q)^-$. The
results are shown in Table \ref{Numtable}. In the Table, we consider
$M_N = 5$ GeV and 50 GeV. For other neutrino masses that obey $M_N \ll
M_W$, such as $M_N = 1$ GeV or 10 GeV, the numbers do not differ much
from those for $M_N = 5$ GeV.

\begin{table}
\begin{center}
\begin{tabular}{ |c|c|c|c|c|  }
\hline
Machine & \multicolumn{2}{|c|}{$\sigma$ (fb): $\ell_1^- \ell_2^-  jj$} &
\multicolumn{2}{|c|}{$\mathcal{N}_{\rm events}$ $(\times 10^{-3})$} \\
 \hline
& $M_N = 5$ GeV & $M_N = 50$ GeV & $M_N = 5$ GeV & $M_N = 50$ GeV \\ 
\hline
	HL-LHC  & 51.7 & 22.3 & 155.1 & 66.9    \\
	HE-LHC  & 98.1 & 42.0 &1471.5 & 630 \\
	FCC-hh  & 323.8 & 136.7  &9714 & 4101  \\ \hline
Machine & \multicolumn{2}{|c|}{$\sigma$ (fb): $\ell_1^+ \ell_2^+  jj$} &
\multicolumn{2}{|c|}{$\mathcal{N}_{\rm events}$ $(\times 10^{-3})$} \\
 \hline
& $M_N = 5$ GeV & $M_N = 50$ GeV & $M_N = 5$ GeV & $M_N = 50$ GeV \\ 
\hline
	HL-LHC &80.0 & 31.9 & 240 & 95.7  \\
	HE-LHC &131.0 & 52.8 & 1965 & 792  \\
	FCC-hh &358.2 & 147.6 & 10746 & 4428  \\
        \hline
\end{tabular}
\end{center}
\caption{\small Predicted cross sections and number of events for $pp
  \to X \ell_1^- \ell_2^- (q'{\bar q})^+$ and $pp \to X \ell_1^+
  \ell_2^+ ({\bar q}'q)^-$. Neutrino masses $M_N = 5$ and 50 GeV are
  considered. Results are given for the HL-LHC ($\sqrt{s} = 14$ TeV,
  peak ${\cal L}_{\rm int} = 3~{\rm ab}^{-1}$), HE-LHC ($\sqrt{s} = 27$ TeV,
  peak ${\cal L}_{\rm int} = 15~{\rm ab}^{-1}$), and FCC-hh ($\sqrt{s} = 100$ TeV,
  peak ${\cal L}_{\rm int} = 30~{\rm ab}^{-1}$).}
\label{Numtable}
\end{table}

We also present in Table \ref{Numtable} the expected number of events,
based on the cross section and integrated luminosity of the machine.
However, that is not necessarily the final answer. The second
ingredient is to look at the $N$ lifetime and see what percentage of
the heavy neutrinos produced in the $W$ decays actually decay in the
detector.  To obtain the number of {\it measurable} events, one must
multiply the expected number of produced events by this percentage.

For a given value of $M_N$, it is straightforward to find the neutrino
lifetime, and to convert this into a distance traveled.  However, the
question of how many neutrinos actually decay in the detector depends
on the size of the detector, and this depends on the particular
experiment. As an example, we note that, in its search for $W^- \to
\ell_1^- \ell_2^- (f'{\bar f})^+$, the CMS Collaboration considered
this question \cite{Sirunyan:2018mtv}. They found that, for $M_N = 10$
GeV, there was essentially no reduction factor, i.e., the percentage of
neutrinos decaying in the detector was close to 100\%. However, for
$M_N = 5$ GeV, the reduction factor was 0.1, while for $M_N = 1$ GeV,
it was $10^{-3}$. Thus, the efficiency of a given experiment for
observing this decay, and measuring $\ACP$, depends on this reduction
factor.

For a given value of $M_N$, one can determine the reduction factor,
and hence the total number of measurable events $N_{\rm events}$.  In
order to turn this into a prediction for the smallest value of
$|\ACP|$ that can be measured at $n\sigma$, the experimental
efficiency must be included. In Refs.~\cite{Khachatryan:2015gha,
  Khachatryan:2016olu}, the CMS Collaboration searched for heavy
Majorana neutrinos at the $\sqrt{s} = 8$ TeV LHC using the final state
$\ell_1^- \ell_2^- j j$.  Including backgrounds, detector efficiency,
etc., their overall efficiency was $\sim 1$\%.

Using an overall efficiency of 1\%, in Table \ref{ACPminTable} we
present the minimum values of $\ACP$ measurable at $3\sigma$ at the
HL-LHC, HE-LHC and FCC-hh. The results are shown for $M_N = 5$ GeV
(with a reduction factor of 0.1), $M_N = 10$ GeV (with no reduction
factor), and $M_N = 50$ GeV (with no reduction factor).

From this Table, we see that, as the LHC increases in energy and
integrated luminosity, smaller and smaller values of $\ACP$ are
measurable.  The most promising results are for $M_N = 10$ GeV, but in
all cases reasonably small values of $\ACP$ can be probed. 

\begin{table}
\begin{center}
\begin{tabular}{ |c|c|c|c|  } \hline
  \multicolumn{4}{|c|}{Minimum $\ACP$ measurable at $3\sigma$} \\
\hline
Machine & $M_N = 5$ GeV & $M_N = 10$ GeV & $M_N = 50$ GeV \\
\hline
HL-LHC & 15.0\% & 4.8\% & 7.4\% \\
HE-LHC & 5.1\% & 1.6\% & 2.5\% \\
FCC-hh & 2.1\% & 0.7\% & 1.0\% \\
\hline
\end{tabular}
\end{center}
\caption{\small Minimum value of $\ACP$ measurable at $3\sigma$ at the
  HL-LHC ($\sqrt{s} = 14$ TeV, peak ${\cal L}_{\rm int} = 3~{\rm ab}^{-1}$),
  HE-LHC ($\sqrt{s} = 27$ TeV, peak ${\cal L}_{\rm int} = 15~{\rm ab}^{-1}$),
  and FCC-hh ($\sqrt{s} = 100$ TeV, peak ${\cal L}_{\rm int} = 30~{\rm
    ab}^{-1}$). Results are given for $M_N = 5$ GeV (reduction factor
  $= 0.1$), $M_N = 10$ GeV (no reduction factor), and $M_N = 50$ GeV
  (no reduction factor).}
\label{ACPminTable}
\end{table}

\newpage

\section{Summary \& Discussion}

Two subjects whose explanation requires physics beyond the SM are
neutrino masses and the baryon asymmetry of the universe. The standard
method for generating tiny neutrino masses is the seesaw mechanism, in
which one introduces three right-handed neutrinos $N_i$. As for the
baryon asymmetry, leptogenesis is often used: CP-violating,
lepton-number-violating processes produce a lepton asymmetry, and this
is converted into a baryon asymmetry through sphaleron processes.
Models that combine these two ideas often involve a quasi-degenerate
pair of heavy neutrinos $N_1$ and $N_2$; leptogenesis arises through
the CP-violating decays of these heavy neutrinos.

Here, an intriguing aspect is that the masses of $N_{1,2}$ can be
small, $O({\rm GeV})$. This has led to suggestions to look for
CP-violating LNV effects in decays of light mesons or $\tau$
leptons. These processes all involve the exchange of a virtual
$W$. However, one can also consider CP-violating LNV decays of real
$W$s at the LHC. Indeed, searches for LNV at the LHC use the decay
$W^\pm \to \ell_1^\pm \ell_2^\pm (q'{\bar q})^\mp$.  In this paper, we
have examined the prospects for measuring CP violation in such decays
at the LHC.

The point is that the decay $W^\pm \to \ell_1^\pm \ell_2^\pm (q'{\bar
  q})^\mp$ arises via $W^{\pm} \to \ell_1^\pm N_i)$, with $N_i \to
\ell_2^\pm W^{*\mp} (\to q'{\bar q})^\mp)$. Here, the $W$-$\ell$-$N_i$
couplings are generated due to the heavy-light neutrino mixing of the
seesaw mechanism. CP violation occurs due to the interference of the
$N_1$ and $N_2$ contributions.

A signal of CP violation would be the measurement of a nonzero
difference in the rates of the decay $W^- \to \ell_1^- \ell_2^-
(q'{\bar q})^+$ and its CP-conjugate. This type of CP asymmetry
requires that the two interfering amplitudes have both CP-odd and
CP-even phase differences.  The CP-odd phase difference is due to
different $W$-$\ell$-$N_1$ and $W$-$\ell$-$N_2$ couplings. A CP-even
phase difference can be generated in two ways, via propagator effects
or oscillations of the heavy neutrino. Both are taken into account in
our study.

Our analysis has two pieces, theory predictions and experimental
prospects. On the theory side, we have computed the expression for the
CP-violating rate asymmetry $\ACP$ [Eqs.~(\ref{ACPsimple}) and
  (\ref{xydefs})]. We consider neutrino masses in the range $5~{\rm
  GeV} \le M_N \le 80~{\rm GeV}$. (The LHC has poor sensitivity to
smaller masses.) For various values of the neutrino mass, we compute
the potential size of $\ACP$. For low masses, e.g., $5~{\rm GeV} \le
M_N \le 20~{\rm GeV}$, we find that (i) the contribution of neutrino
oscillations to the CP-even phase is much suppressed compared to that
from propagator effects, and (ii) $\ACP$ can be large, $\gsim
0.9$. For large masses, e.g., $M_N \ge 60$ GeV, the contribution of
neutrino oscillations to the CP-even phase becomes important, but has
the effect of reducing the CP asymmetry, $\ACP \le 0.6$.

On the experimental side, we want to determine the smallest value of
$\ACP$ that can be measured at $3\sigma$ at the LHC. This depends on
the number of observed events, and we use {\tt MadGraph} to find this
for three versions of the LHC: (i) the high-luminosity LHC (HL-LHC,
$\sqrt{s} = 14$ TeV), (ii) the high-energy LHC (HE-LHC, $\sqrt{s} =
27$ TeV), (iii) the future circular collider (FCC-hh, $\sqrt{s} = 100$
TeV). We assume an experimental efficiency of 1\%
\cite{Khachatryan:2015gha, Khachatryan:2016olu}. The one input
required is the size of the heavy-light neutrino mixing parameter
$|B_{\ell_1 N_1}|^2$. Taking into account the present experimental
constraints, in our analysis we take $|B_{\ell_1 N_1}|^2 = 10^{-5}$.

We find that, while the minimum value of $\ACP$ measurable at the LHC
depends on the neutrino mass $M_N$, smaller and smaller values of
$\ACP$ can be measured as the LHC increases in energy and integrated
luminosity. The most promising result is for the FCC-hh with $M_N =
10$ GeV, where $\ACP = O(1\%)$ is measurable. But even for the worst
case, the HL-LHC with $M_N = 5$ GeV, a reasonably small value of $\ACP
= O(10\%)$ can be measured.

The point to take away from all of this is the following.  The simple
observation of the LNV decay $W^\pm \to \ell_1^\pm \ell_2^\pm (q'{\bar
  q})^\mp$ would itself be very exciting. But the next step would then
be to try to understand the underlying new physics. If a CP asymmetry
in this decay were measured, it would tell us that (at least) two
amplitudes contribute to the decay, with different CP-odd and CP-even
phases, and would hint at a possible connection with leptogenesis
models.

\bigskip
{\bf Acknowledgments}: This work was financially supported by NSERC of
Canada.

\end{document}